%% file: main.tex
\documentclass[preprint]{article}
\usepackage[letterpaper, margin=1in]{geometry}
\usepackage{graphicx}
\usepackage{newtxtext}
\usepackage{newtxmath}
\usepackage{natbib}
\usepackage{hyperref}
\usepackage{subcaption}
\usepackage{bbold}
\usepackage{bm}
\usepackage{authblk}

\hypersetup{
    colorlinks = true,
    urlcolor   = blue,
    citecolor  = black,
}

\newcommand{\RomanNumeralCaps}[1]
\linenumbers

\DeclareMathOperator*{\argmin}{arg\,min}

\def\bi#1{\textbf{#1}}
\def\mx#1{#1}

\title{Learning fluid physics from highly turbulent data using sparse physics-informed discovery of empirical relations (SPIDER)}

\author[1]{Daniel R. Gurevich}
\author[2]{Matthew R. Golden}
\author[2]{Patrick A. K. Reinbold}
\author[2]{Roman O. Grigoriev}

\affil[1]{Program in Applied and Computational Mathematics, Princeton University, Princeton, NJ 08544}
\affil[2]{{School of Physics, Georgia Institute of Technology, Atlanta, GA 30332}}

\begin{document}
\maketitle

\begin{abstract}
We show how a complete mathematical model of a physical process can be learned directly from data via a hybrid approach combining three simple and general ingredients: physical assumptions of smoothness, locality, and symmetry, a weak formulation of differential equations, and sparse regression. To illustrate this, we extract a complete system of governing equations of fluid dynamics---the Navier-Stokes equation, the continuity equation, and the boundary conditions---as well as the pressure-Poisson and energy equations, from numerical data describing a highly turbulent channel flow in three dimensions. Whether they represent known or unknown physics, relations discovered using this approach take the familiar form of partial differential equations, which are easily interpretable and readily provide information about the relative importance of different physical effects. The proposed approach offers insight into the quality of the data, serving as a useful diagnostic tool. It is also remarkably robust, yielding accurate results for very high noise levels, and should thus be well-suited for analysis of experimental data.
\end{abstract}

%\begin{keywords}
%Authors should not enter keywords on the manuscript, as these must be chosen by the author during the online submission process and will then be added during the typesetting process (see \href{https://www.cambridge.org/core/journals/journal-of-fluid-mechanics/information/list-of-keywords}{Keyword PDF} for the full list).  Other classifications will be added at the same time.
%\end{keywords}

%{\bi MSC Codes }  {\it(Optional)} Please enter your MSC Codes here

% Changes to make:
% i. add a couple of references and explain difference from past work more explicitly in intro
% iv. mention improvements over SINDy after explanation of SPIDER

\section{Introduction}

Physical theories are traditionally constructed in an iterative manner. At each step, discrepancies between predictions and existing experimental observations are used to improve the theory, making it more general and accurate. These improvements are usually instructed and constrained by first principles, including both general and domain knowledge. After this, new predictions are made and new experiments are designed to test these predictions, closing the loop. Humans play a key role in all aspects of this traditional procedure and can become a weak link when the amount of data becomes overwhelming or the patterns in the data are too complex. 

Recent advances in machine learning have started to change the scientific paradigm guiding the construction of physical theories by gradually taking humans out of the loop. For low-dimensional systems, physical relations in the form of algebraic and even differential equations can be constructed using symbolic regression directly from experimental data without using any physical intuition \citep{crutchfield1987,bongard2007,schmidt2009}. For high-dimensional systems such as fluid flows, purely data-driven approaches often become intractable, and some physical intuition becomes necessary to guide the process \citep{karpatne2017}. 

The question is therefore what physical considerations can and should be used to constrain the problem sufficiently for the data-driven analysis to become tractable while leaving enough freedom to enable identification of physically meaningful relationships. Among the most general and least restrictive physical constraints are {\it smoothness}, {\it locality}, and the {\it relevant symmetries}. In fact, some or all of these constraints have been implicitly assumed in most efforts to identify {\it evolution equations} via some form of regression from synthetic data \citep{bar1999,xu2008,rudy2017,schaeffer2017,reinbold2020} or experimental data \citep{reinbold2021}. However, evolution equations are just one type of a relation that may be required to fully describe a physical system. Other examples include {\it constraints}, such as the divergence-free condition for the velocity field representing mass conservation for an incompressible fluid or the curl-free condition for the electric field in electrostatics, as well as {\it boundary conditions}. Previous studies have largely ignored the problem of identifying these equally important classes of relations for high-dimensional systems.

Sparse linear regression has so far proven to be the most versatile and robust approach for equation inference. Its original implementations, such as Sparse Identification of Nonlinear Dynamics (SINDy) algorithm \citep{brunton2016}, were aimed at discovering evolution equations. Generalizations of this algorithm such as SINDy-PI \citep{kaheman2020}, which find sparse solutions to a collection of inhomogeneous linear systems, can be used to discover other types of relations as well. 
%A more computationally efficient and straightforward approach is to solve a single {\it homogeneous} linear system \citep{golden2023}. 
A number of alternatives for nonlinear regression aimed at inference of PDEs have been proposed as well. These include Gaussian processes \citep{raissi2018}, gene expression programming (GEP) \citep{ferreira2001, xing2022,ma2022} and several neural network-based approaches such as Equation Learner (EQL) \citep{martius2016,sahoo2018}, Neural Symbolic Regression that Scales (NeSymReS) \citep{biggio2021}, PDE-LEARN \citep{stephany2022}, and PDE-Net \citep{long2018}. While most of these approaches have been validated by reconstructing canonical PDEs or {\it known} governing equations, their potential for discovering previously {\it unknown} physics remains unclear, especially for spatially extended systems in more than one spatial dimension.

All of the above approaches suffer from inherent sensitivity to noise in the data which is amplified by spatial and/or temporal derivatives that appear in any physical relation desctibed by a partial differential equation (PDE). When the strong form of PDEs is used, it becomes difficult or even impossible to correctly identify governing equations involving higher-order derivatives for noise levels as low as a few percent \citep{rudy2017,raissi2018,raissi2019}. This sensitivity can be addressed by using the weak form of governing equations \citep{gurevich2019}, as illustrated by its successful application in equation inference approaches employing both linear regression \citep{reinbold2020,messenger2021,alves2022} and nonlinear regression \citep{stephafny2023}. Weak formulation was also found to be useful in problems involving latent variables \citep{reinbold2021} and unreliable or missing data \citep{golden2023}.

The success of any approach to equation inference ultimately depends on the availability of a sufficiently rich function library (or, more typically, multiple libraries) which define the search space for one or more parsimonious relations describing the data. With rare exceptions, these libraries have previously been constructed in a largely {\em ad hoc} manner, either with little regard for the specifics of the physical problem or, alternatively, relying too much on the presumed-to-be-known physics. In this article, we describe a flexible and general data-driven approach for identifying a complete mathematical description of a physical system, including relevant boundary conditions, which we call Sparse Physics-Informed Discovery of Empirical Relations (SPIDER). Unlike SINDy and its variants, SPIDER is more than a linear regression algorithm: it is based on a {\it systematic} procedure for library generation informed by the symmetries of the system. 
% Talk about differences with SA paper
%\textcolor{red}{The application of SPIDER presented in this manuscript has several differences with the experimental application \citet{golden2023}. Namely, the data is 3+1 dimensional and on a nonuniform spatial grid. Since the flow is not in the Stokes limit, the libraries are also allowed to contain nonlinear terms in the flow velocity. Further, the present implementation discovers the boundary conditions of the flow.} 
We illustrate SPIDER by discovering the evolution equations, constraints, and boundary conditions governing the flow of an incompressible Newtonian fluid from noisy numerical data using only very mild constraints which require no detailed knowledge of the physics.
The implementation of SPIDER described here is publicly available at \url{https://github.com/sibirica/SPIDER_channelflow}.

The paper is organized as follows. Our hybrid equation inference approach is introduced and illustrated using an example of data representing numerical simulation of a highly turbulent flow in Section \ref{sec:spider}. The results are discussed in Section \ref{sec:disc}, and our conclusions are presented in Section \ref{sec:concl}.

\section{Sparse physics-informed discovery of empirical relations}
\label{sec:spider}

\begin{figure}
\centering
\includegraphics[width=0.45\columnwidth]{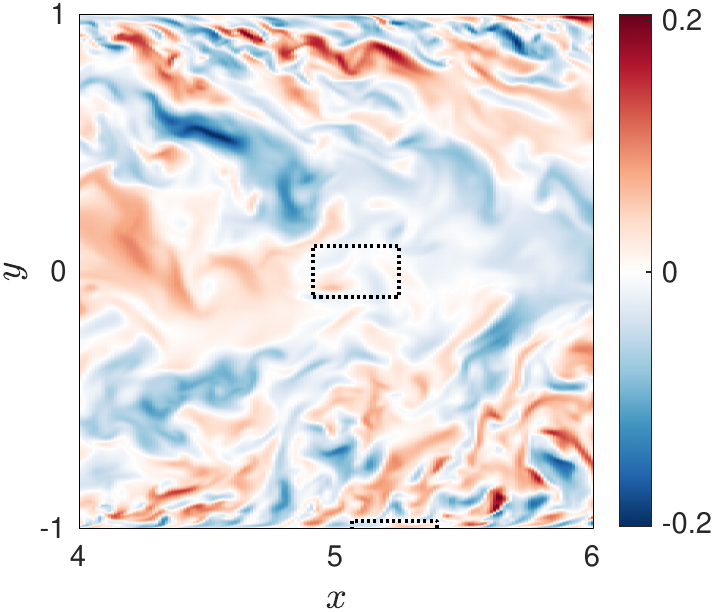}
\caption{Snapshot of the velocity component $u_z$ in a $z={\rm const}$ plane over a portion of the entire computational domain. Sample integration domains (shown as dotted boxes) near the edge of the channel are much narrower than those in the middle due to the non-uniform grid spacing in the $y$ direction.
}
\label{fig:domains}
\end{figure}

It is well known that, in order for a data-driven approach to identify a sufficiently general mathematical model, the data must exhibit enough variation to sample the state space of the physical problem \citep{schaeffer2018}. Here, this is accomplished by using the numerical solution of a high-Reynolds number flow through a rectangular channel from the Johns Hopkins University turbulence database (\url{http://turbulence.pha.jhu.edu/Channel_Flow.aspx}). %\citep{jhutdb}. 
The data set includes the flow velocity ${\bi u}$ and pressure $p$ fully resolved in space and time. The channel dimensions are $L_x\times L_y\times L_z\times L_t=8\pi\times 2\times 3\pi\times 26$ (in nondimensional units) and the data are stored on a spatiotemporal grid of size $2048\times 512\times 1536\times 4000$. The (nondimensional) viscosity is $\nu=5\times 10^{-5}$ and the corresponding friction Reynolds number is $Re_\tau \sim 10^3$. A representative snapshot of the data is shown in Figure \ref{fig:domains}. 

The immense size of the entire data set comprising $2.6\times 10^{13}$ ``measurements'' illustrates the challenges faced by a purely data-driven approach. The locality property radically reduces the number of possible functional relations between measurements by constraining these to a small spatiotemporal neighborhood of a given point. In particular, for smooth continuous fields, such functional relations have to be expressed in terms of their values and partial derivatives. For systems that are invariant with respect to spatial and temporal translation,
a functional relation can be expressed in the form of a Volterra series
\begin{equation}\label{eq:volterra}
\sum_{n=1}^N c_n {\bi f}_n = 0,
\end{equation}
where $c_n$ are coefficients and ${\bi f}_n$ are products of the fields and their partial derivatives. For systems with translational symmetry in space and time, the most general relations of this type are nonlinear PDEs with constant coefficients. Most prior work focused on evolution equations, which are special cases of \eqref{eq:volterra} where $c_1=1$ and ${\bi f}_1$ is the first-order temporal derivative of one of the fields. Other special cases include differential equations that do not involve temporal derivatives and algebraic relations between the fields that involve no derivatives at all, which have largely been ignored by the machine learning literature in the context of spatially extended systems.

Our aim here is to identify a parsimonious mathematical model of the flow in the form of a system of PDEs, along with appropriate boundary conditions, directly from data representing the velocity and pressure fields, ${\bi u}$ and $p$. The key observation here is that the form of the functional relations \eqref{eq:volterra} can be restricted sufficiently using the rotational symmetry constraint. All terms ${\bi f}_n$ have to transform in the same way under rotations and reflections, with the transformation rule corresponding to a particular representation of the orthogonal symmetry group O(3). For non-relativistic systems, the symmetry group involves rotations about any axis in three-dimensional space and reflections across any plane, with the representations corresponding to tensors of various ranks. Here we will restrict our attention to the two lowest rank tensors, i.e., scalars and vectors, although the same approach trivially extends to tensors of any rank \citep{golden2023}.  

\subsection{Learning evolution equations and constraints}

%\subsubsection{Constructing model libraries} 
The functional form of the mathematical model will always depend on the choice of the variables. The best choice may not be obvious, and this is where relevant domain knowledge is extremely helpful. In the present problem, we will assume that the variables are the pressure field $p$ and the velocity field ${\bf u}$ and that both variables are fully observed. The pressure is a scalar and the velocity is a vector. The differential operators $\partial_t$ and $\nabla$ transform as a scalar and a vector, respectively. Using these four objects, we can construct tensors of any rank using tensor products and contractions \citep{golden2023}. For instance, the terms ${\bi u}$, $\partial_t{\bi u}$, and $\nabla p$ all transform as vectors. To illustrate the procedure, we will include {\it all} possible terms ${\bi f}_n$ up to cubic in $p$, ${\bi u}$, $\partial_t$, and/or $\nabla$ that can be constructed from the data and its derivatives, yielding a scalar library
\begin{equation}
\begin{aligned}
\mathcal{L}_0=\{1,p,\nabla\cdot{\bi u},\partial_t p,p^2,{\bi u}^2, p^3, {\bi u}\cdot\nabla p,\nabla^2p,
p\partial_tp,\partial^2_tp,p(\nabla\cdot {\bi u}),{\bi u}^2p,{\bi u}\cdot\partial_t{\bi u}\}
\label{eq:rank0}
\end{aligned}
\end{equation}
and a vector library
\begin{equation}
\begin{aligned}
\mathcal{L}_1=\{&{\bi u},\partial_t{\bi u},\nabla p,p{\bi u},({\bi u}\cdot\nabla){\bi u},\nabla^2{\bi u},\partial^2_t{\bi u},u^2{\bi u},p^2{\bi u},\\
&\partial_t \nabla p, p\nabla p,{\bi u}(\nabla\cdot{\bi u}),
{\bi u}\cdot(\nabla{\bi u}),%\nabla({\bi u^2}),
\nabla(\nabla\cdot{\bi u}),p \partial_t {\bi u},{\bi u} \partial_t p\}.
\label{eq:rank1}
\end{aligned}
\end{equation}
These two libraries, together with the relation \eqref{eq:volterra}, will form the search space containing all of the candidate relations describing the fluid physics in the bulk. Note that scalars and vectors (i.e., rank-0 and rank-1 tensors) are {\it irreducible} representations of the symmetry group O(3). This is not the case for rank-2 tensors, for instance, which can be broken into three different irreducible representations corresponding to the symmetric traceless component, antisymmetric component, and the trace. Similarly, scalars and pseudoscalars, or vectors and pseudovectors,  belong to different irreducible representations of O(3). Reflection covariance can be used to exclude pseudoscalars and pseudovectors such as ${\bf u} \cdot (\nabla \times {\bf u})$ and $\nabla\times{\bi u}$ from the scalar and vector libraries.

%where $\nabla {\bi u}$ is the velocity gradient tensor and $:$ denotes the double dot product. 
It should be emphasized that no domain knowledge specific to the system, aside from the symmetry (rotational and translational) and the choice of variables, has been used in constructing these libraries. For instance, it is not necessary to know that ${\bi u}$ and $p$ represent the velocity and pressure of a fluid. This is in direct contrast to most prior studies \citep{raissi2018,reinbold2020,reinbold2021,messenger2021,ma2022} that used model libraries directly inspired by first-principles analysis of the fluid flows considered there.

It is also useful to put the very modest size of libraries $\mathcal{L}_0$ and $\mathcal{L}_1$ in perspective. In order to identify the evolution equation for the vorticity $\omega=\nabla\times{\bi u}$, \citet{rudy2017} used a library analogous to $\mathcal{L}_1$ that was constructed using a brute-force approach ignoring the symmetries of the problem. The terms that were chosen by the authors 
%(not in a fully systematic way) 
included $\partial_t\omega$ as well as ``polynomial terms of vorticity and all velocity components up to second degree, multiplied by derivatives of the vorticity up to second order,'' yielding a set of $N=1+(1+2d+d(d-1)/2)^2$ terms in $d$ spatial dimensions. For the three-dimensional geometry considered here, the corresponding library would contain 101 distinct terms, almost an order of magnitude more than what is included in our more physically comprehensive library $\mathcal{L}_1$, which was constructed using symmetry constraints. In fact, incorporating knowledge of the Galilean invariance in our system would have allowed for an even more compact library to be used without loss of expressivity.

\subsubsection{The effect of noise}

Two common scenarios where equation inference would be of particular value are when the data are generated experimentally \citep{reinbold2021,joshi2022,golden2023} or when the data represent coarse-graining of the results of direct numerical simulation. An example of the latter is fully kinetic simulations of plasma used to obtain a hydrodynamic description \citep{alves2022}. In both instances, the data will inevitably be noisy, e.g., due to measurement inaccuracies in experiment or fluctuations of the computed coarse-grained fields. To investigate the effects of noise, in addition to the original simulation data downloaded from the turbulence database, we also used synthetic data with varying levels of additive uniform noise. Specifically, we define the noisy data $f_\sigma = f + \sigma \xi_f s_f,$ where $f\in\{p,u_x,u_y,u_z\}$ are the hydrodynamic fields, $\sigma$ is the noise level, and $\xi_f$ is noise independently sampled from the uniform distribution over $[-1,1]$ at each spacetime point.
%$p_{\sigma} = p + \sigma N_1 s_p$ and ${\bi u}_{\sigma} = {\bi u} + \sigma\,\text{diag}\{{\bi N}_2\}{\bi s}_u$, where $\sigma$ is the noise level. We take $N_1$ and the three components of ${\bi N}_2$ to be independently sampled from the uniform distribution over $[-1,1]$ at each point in space and time, and $s_p$ and ${\bi s}_u$ are the sample standard deviations of the pressure and each component of the flow velocity computed across the original dataset. 

Parsimonious scalar and vector relations describing velocity and pressure data can be identified by performing sparse regression using the libraries $\mathcal{L}_0$ and $\mathcal{L}_1$, respectively. In the strong form, the terms involving higher-order derivatives, such as $\nabla^2p$ and $\nabla^2{\bi u}$, will be extremely sensitive to noise \citep{rudy2017,reinbold2019}. 
To make the regression more robust, we use the weak form of both PDEs following the approach introduced in our earlier work \citep{gurevich2019}. 
Specifically, we multiply each equation by a smooth weight function $w_j({\bi x},t)$ and then integrate it over a rectangular spatiotemporal domain $\Omega_i$ of size $H_x\times H_y\times H_z\times H_t$. 
All side-lengths of $\Omega_i$ are fixed to $H_i=32$ grid points of the numerical grid; this size roughly corresponds to the characteristic length and time scales of the flow field (cf. Figure \ref{fig:domains}). 

\begin{figure}[h]
    \centering
    \includegraphics[width=0.50\textwidth]{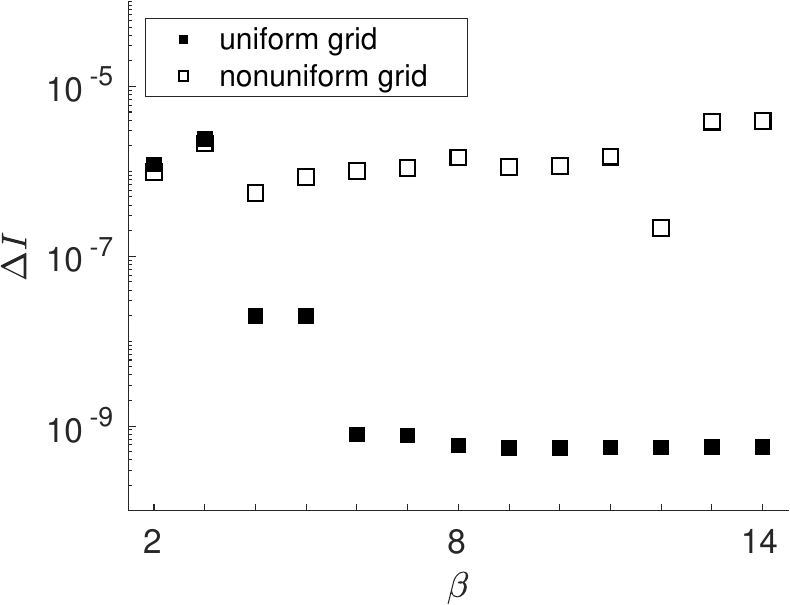}
    
    \caption{
    Numerical error $\Delta I$ of the integral $I_\beta = \int_{-1}^1 \cos(2\pi y) (1-y^2)^\beta \, dy$ evaluated using the trapezoid rule on a 32-point grid. The results are shown for the uniform grid (solid line) and the nonuniform grid (dashed line) representing the $y$-coordinates of points taken from the middle of the channel. %The numerical integral $I_{\beta, \textrm{trap}}$ is computed by trapezoidal quadratures. 
    }
    \label{fig:integration_error}
\end{figure}

The derivatives are shifted from the data (${\bi u}$ and $p$) onto the weight functions $w$ whenever possible via integration by parts, after which the integrals are evaluated numerically using trapezoidal quadratures. 
For the scalar library $\mathcal{L}_0$, we use scalar weight functions of the form
\begin{align} \label{eq:wf}
&w({\bi x},t) = \tilde{w}(\bar{x})\tilde{w}(\bar{y})\tilde{w}(\bar{z})\tilde{w}(\bar{t}), \nonumber\\
&\tilde{w}(s) = (1-s^2)^\beta,
\end{align}
where the bar denotes nondimensionalization using the order-preserving affine map $[x_{\textrm{min}},x_{\textrm{max}}] \to [-1,1]$. 
In the case of the vector library $\mathcal{L}_1$, we instead use vector weight functions $w({\bi x}, t){\bi e}_k$ aligned along each of the coordinate axes.
%Here we chose $l=0$ or $1$ for each of the coordinate directions, yielding $2^{d+1}$ different weight functions for 
%a $(d+1)$-dimensional integration domain. 
%an integration domain with one temporal and $d$ spatial dimensions.
Note that, for the term libraries considered in this problem which involve a Laplacian of $p$ or ${\bf u}$, we should have $\beta\ge 2$, as this allows discarding the boundary terms generated during integration by parts. We set $\beta = 8$ in our analysis since this choice (i) ensures that all the boundary terms vanish and (ii) maximizes the accuracy of numerical quadrature along the uniformly gridded dimensions \citep{gurevich2019}. The nonuniform grid in the $y$-direction will control the error of the quadrature and increasing $\beta$ further has no benefit. This is illustrated in Figure \ref{fig:integration_error}, which shows how the quadrature error scales with $\beta$ for a test function $\cos(2\pi y)$. %To integrate \eqref{eq:rank1}, we instead use vector weight functions ${\bi w}^i_j({\bi x},t) = w_j({\bi x},t)\hat{\b i}$ for $i=1,\cdots,d$ varying over the indices of the spatial dimensions, resulting in a total of $d\,2^{d+1}$ different weight functions. 
The error decreases quickly with increasing $\beta$ for the uniform grid, plateauing at $\beta\ge 8$. In contrast, the error is found to be almost independent of $\beta$ for the nonuniform grid.

By repeating this procedure for a number of integration domains $\Omega_i$ contained within the full dataset, we construct a feature matrix $\mx{Q} = [{\bi q}_1 \dots {\bi q}_N]$  whose the columns ${\bi q}_n$ correspond to the various terms in either $\mathcal{L}_0$ or $\mathcal{L}_1$. For instance, for the scalar library $\mathcal{L}_0$,
\begin{equation}
    Q_{ij} = \frac{1}{V_i \, S_j}\int_{\Omega_i} w({\bi x},t) f_j\, d^3{\bi x}\, dt,
\end{equation}
where $V_i$ is the integral of $w$ over $\Omega_i$ and $S_j = S[f_j]$ is the 
%physics-informed 
scale of the library term $f_j$ in dimensional units. {We estimate the scales of all library terms using the characteristic time scales $T_{\bi u},T_p$, length scales $L_{\bi u}, L_p$, mean values $\mu_{\bi u}, \mu_p$, and standard deviations $\sigma_{\bi u}, \sigma_p$ of the velocity and pressure fields across the dataset. For instance, for a vector field ${\bi f}$, the length and time scales can be estimated from the magnitudes of its derivatives:
\begin{align}
    & T_{{\bi f}} \equiv \frac{\sigma_{\bi f}}{ \sqrt{ \langle \partial_t {\bi f} \cdot \partial_t {\bi f} \rangle} },\\
    & L_{{\bi f}} \equiv \frac{\sigma_{\bi f}}{ \sqrt{ \langle (\nabla {\bi f}) \cdot (\nabla {\bi f}) \rangle}},
\end{align}
where the dot products in the denominators are taken over all the indices. We then use the heuristic that the scale of a library term is the product of the scales of its factors: $S[{\bi f \bi g}] = S[{\bi f}] S[{\bi g}]$, and for $k \geq 1$
\begin{align}
    & S[p] = \mu_p, && S[{\bi u}] = \mu_{\bi u},\\
    & S[\nabla^k p] = L_p^{-k} \sigma_p, && S[{\nabla^k {\bi u}}] = L_{\bi u}^{-k} \sigma_{\bi u},\\
    & S[{\partial_t^k p}] = T_p^{-k}, \sigma_p && S[{\partial_t^k {\bi u}}] = T_{\bi u}^{-k} \sigma_{\bi u}.
\end{align}
%For variables that do not appear under a derivative such as ${\bi u}^2$, we use the mean $\langle \cdot \rangle$ to determine the scale: $S = \langle |{\bi u}| \rangle^2$. For differentiated variables, the scale is estimated in terms of the standard deviation $[ \cdot ]$ of the field and the inverse of the corresponding length scale or time scale \textcolor{red}{for the derivative applied}. For instance, the scale of $({\bi u} \cdot \nabla) {\bi u}$ is $S = \langle |\bi u| \rangle L^{-1}_u [|\bi u|]$. The length and time scales of variables can be estimated by means of evaluating spatiotemporal gradients. All scales used are products of means, standard deviations, and spatiotemporal scales. 
Note that the mass scale does not appear explicitly, as the data are given in units in which the density $\rho=1$.} The nondimensionalization procedure ensures the magnitudes of all columns are comparable, which can dramatically improve the accuracy and robustness of regression. The problem of determining the unknown coefficients ${\bi c}=[c_1,\cdots,c_N]^\top$ is cast as the solution of an overdetermined linear system of the form
\begin{align}
    \mx{Q}{\bi c} = 0.
    \label{eq:lin}
\end{align}
%where $\mx{Q} = [{\bi q}_1 \dots {\bi q}_N]$ and the columns ${\bi q}_n$ correspond to the integrals of the terms ${\bi f}_n$ in \eqref{eq:volterra}. 
In this article, we sample $\Omega_i$ from a $64^4$-point region of the data lying either in the middle (i.e., the symmetry plane of the flow) or at the edge of the channel (see Table \ref{tab:locations}). Examples of both types of domains are shown in Figure \ref{fig:domains}.
%Allowing the domains to have a $50\%$ overlap by volume yields $M=256$ linear equations on the coefficients of $\mathcal{L}_0$ and $3M=768$ linear equations on the coefficients of $\mathcal{L}_1$.
We use 256 randomly sampled subdomains $\Omega_i$ with $32^4$ gridpoints. This yields $M=256$ linear equations on the coefficients of $\mathcal{L}_0$ and $3M=768$ linear equations on the coefficients of $\mathcal{L}_1$.

%\begin{table}
%\centering
%\begin{tabular}{*{5}{c}}
% & $H_x$ & $H_y$ & $H_z$ & $H_t$ \\ \hline
%edge & 1/3 (27) & 1/50 (34) & 1/5 (33) & 1/4 (38) %\\ 
%center & 1/3 (27) & 1/5 (32) & 1/5 (33) & 1/4 (38) \\ \hline
%\end{tabular}
%\caption{Dimensions of the integration domains $\Omega_k$ in physical units and grid points (in parentheses).}
%\label{tab:domain}
%\end{table}

\subsubsection{Selection of parsimonious relations} 

Note that the linear system \eqref{eq:lin} is {\it homogeneous} and treats all terms in the library on equal terms. This is in contrast to SINDy \citep{brunton2016} and its variants \citep{messenger2021} that solve an {\it inhomogeneous} linear system, or many such systems in the case of SINDy-PI \citep{kaheman2020}.
Its solutions have a degree of freedom corresponding to the normalization of ${\bi c}$, which can be eliminated by arbitrarily setting one of the coefficients, say $c_1$, to unity, as done in SINDy, or by fixing the norm of ${\bi c}$ as in this study. The solutions of a constrained least squares problem
\begin{equation}
{\bi c}=\argmin_{\|{\bi c}\|=1}\|\mx{Q}{\bi c}\|
\label{eq:linn}
\end{equation}
are given by the right singular vector of $\mx{Q}$ corresponding to the smallest singular value.
It is worth noting that, when multiple singular values of $\mx{Q}$ are small, there may be several ``good'' independent solutions for ${\bi c}$ representing different dominant balances. This is a less restrictive approach compared to SINDy and allows a broader class of functional relations to be identified. It is also more computationally efficient than SINDy-PI, which aims to address the same limitation.

In order to obtain a parsimonious physical relation, we must find a sparse coefficient vector ${\bi c}^*$ such that the residual $\|\mx{Q}{\bi c}^*\|$ is comparable to the residual $\|\mx{Q}{\bi c}\|$ with dense ${\bi c}$ given by \eqref{eq:linn}. 
The identified relations either contain a single term or several terms. If the matrix $Q$ has been properly nondimensionalized, single-term relations will correspond to columns with small norms. We will use the heuristic that ${\bf f}_j = 0$ is a valid single-term relation if 
$\|{\bf q}_j\| \ll \sqrt{M}$, 
%$\|{\bf q}_j\| < 10^{-3} \sqrt{M}$, 
where $M$ is the number of integration domains. In particular, for the data without added noise, the scalar library $\mathcal{L}_0$ is found to contain terms with $\| {\bf q}_j \| \approx 10^{-6} \sqrt{M}$, which correspond to the incompressibility condition
\begin{align}\label{eq:divu}
    \nabla\cdot{\bi u}=0
\end{align}
and its trivial corollary $p \nabla \cdot {\bf u} = 0$. Both single-term relations are found using data from the middle of the channel as well as data near the boundary. The single-term relation heuristic crucially relies on proper nondimensionalization such that velocity gradients are $O(1)$. A more general approach is to compare with the characteristic size of the uncontracted tensor, which in this case is the rate of strain $\nabla_i u_j$.

The libraries should be pruned to prevent redundancy once any single-term relation is identified. For instance, once the incompressibility condition has been identified, the terms $\nabla \cdot {\bi u}$ and $p(\nabla \cdot {\bi u})$ can be removed from the scalar library $\mathcal{L}_0$, while ${\bi u}( \nabla \cdot {\bi u})$ and $\nabla( \nabla \cdot {\bi u})$ can be removed from the vector library $\mathcal{L}_1$. 
%$(\nabla \cdot {\bi u})^2$ can be removed from \eqref{eq:augment_rank0} \textcolor{red}{\bf Is it ok that we reference an equation not yet seen?} ,
In contrast, prior sparsification algorithms such as SINDy \citep{brunton2016}, implicit SINDy \citep{mangan2016}, and SINDy-PI \citep{kaheman2020} are unable to identify single-term relations, as these are not examined separately.  Note that direct identification of single-term relations is both more robust and more computationally efficient than identification through regression.

%For this reason, we start by computing the magnitude $\|{\bi q}_n\|$ of each term, which equals the residual of the corresponding single-term model. Comparing $\|{\bi q}_n\|$ with the residual $r$ for the identified multi-term model allows determination of whether the parsimonious relation should include one term (if $\min_n\|{\bi q}_n\|$ is below a chosen threshold and $r$ is not) or several terms (otherwise). With this enhancement, SPIDER correctly identified the continuity equation \eqref{eq:divu} from the scalar libraries \eqref{eq:rank0} and \eqref{eq:augment_rank0} for data from either the edge or the center of the channel, across all replications and for all noise levels up to the maximum, $\sigma = 50\%$. In contrast, prior sparsification algorithms such as SINDy \citep{brunton2016} and implicit SINDy \citep{mangan2016} are unable to identify such single-term models, as these are not examined separately.

Once the library has been pruned, multiple-term relations can %usually
be identified by an iterative greedy algorithm. At each iteration, we use the singular value decomposition of $\mx{Q}^{(N)} = [{\bi q}_1 \dots {\bi q}_N]$ to find ${\bi c}^{(N)}$ as described previously. We also compute the residual $r^{(N)} = \|\mx{Q}^{(N)} {\bi c}^{(N)}\|$. Next, we consider all of the candidate relations formed by dropping one of the terms and eliminating the corresponding column from $\mx{Q}^{(N)}$. We select the candidate relation with $N-1$ terms that achieves the smallest residual and then repeat until only one term remains. This yields a sequence of increasingly sparse relations described by $N$-dimensional coefficient vectors ${\bi c}^{(N)}$, forming an approximately Pareto-optimal set \citep{miettinen2012}. Note that the use of the absolute residual $r$ guarantees that the residual is a monotonic function of the number of terms $N$, which is not the case for the relative residual $\eta=r/\max_n\|c_n{\bf q}_n\|$ used by \citet{reinbold2021} and \citet{golden2023}.

There are many reasonable ways to select a final relation from this sequence based on the trade-off between their parsimony (i.e., number of terms $N$) and accuracy, quantified by the residuals $r^{(N)}$. For instance, one might select the simplest relation which achieves a relative residual of less than, say, $1\%$ or the relation for which discarding a single term results in the largest relative increase in the residual. In this article, we follow \citet{gurevich2019}: specifically, we choose the relation described by the coefficient vector ${\bi c}^{(N)}$ where $N = \max\{n : r^{(n)}/r^{(n-1)} > \gamma\}$, where the parameter $\gamma = 1.25$ was selected empirically.

After a sparse relation corresponding to $c^*$ has been identified, one may search for additional sparse relations contained within the same library by further pruning, e.g., discarding one of the terms (usually the largest one) in the previously identified relation from the library. 
%The sparse regression process can then be continued with this pruned library. %For instance, in the scalar case, %\eqref{eq:augment_rank0}, 
%two further relations are identified in this way,
%after removing $\nabla \cdot {\bi u}$ and $p \nabla \cdot {\bi u}$ from the library, 
For instance, we have used this procedure to successfully identify the pressure-Poisson equation and the energy equation from the scalar library expanded to include relevant terms, as discussed below. 

%\begin{figure*}
%\centering
%\subfloat[]{\includegraphics[width=0.47\textwidth]{PNAS/energy_equation.png}}
%\hfill
%\subfloat[]{\includegraphics[width=0.47\textwidth]{PNAS/pressure_poisson.png}}
%\caption{Dependence of the residual $\|Q{\bi c}\|$ on the number of terms $N$ retained in the vector relation \eqref{eq:rank1} for (a) noiseless data and (b) data with 50\% noise. Black (white) squares represent data collected near the edge (in the middle) of the channel. The models selected by the greedy algorithm in each case are shown by the arrows.}
%\label{fig:scalar_residuals}
%\end{figure*}

\subsubsection{Identified equations and robustness to noise} 

%\begin{table}[h]
  %  \centering
  %  \begin{tabular}{cccc}
  %      $\sigma$ &  identified model & & $\|Qc\|$ \\
 %       \hline \\ [-2ex]
 %        5\%   & Navier-Stokes & $\partial_t {\bf u} + \nabla \cdot ({\bf u u}) + \nabla p - \nu \nabla^2 {\bf u} = 0$ & $1.1\times 10^{-3}$ \\
%         50\%  & Euler         & $\partial_t {\bf u} + \nabla \cdot ({\bf u u}) + \nabla p = 0$ & $1.1\times 10^{-2}$ \\
        % 150\% & inviscid Burgers' & $\partial_t {\bf u} + \nabla \cdot ({\bf u u})= 0$ & $3.1 \times 10^{-2}$
%    \end{tabular}
%    \caption{\textcolor{red}{[Don't really need this table here anymore - consider putting it in discussion instead]} Vector models identified when increasing levels of noise are added to data in the middle of the channel.}
%    \label{tab:noisy_vector}
%\end{table}

\begin{figure*}
\centering
%\subfloat[]{\includegraphics[width=0.33\textwidth]{figures2023/div.pdf}}
%\hfill
\captionsetup[subfigure]{oneside,margin={1.1cm,0cm}} %move label to right to align with N
\subfloat[]{\includegraphics[height=80pt]{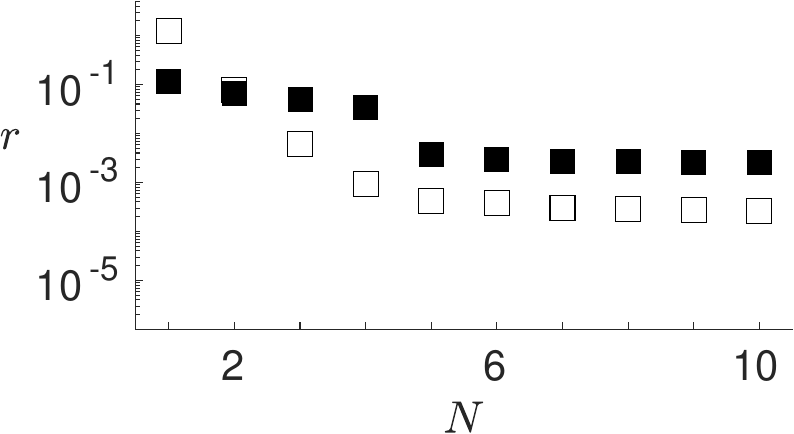}}
\hspace{2mm}
\captionsetup[subfigure]{oneside,margin={0.8cm,0cm}} %reset for figures (b) and (c)
\subfloat[]{\includegraphics[trim={0cm 0 0 0},clip,height=80pt]{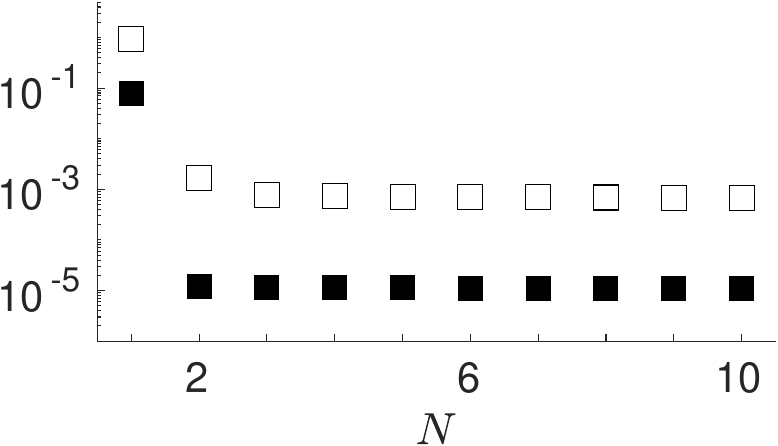}}
\hspace{2mm}
\captionsetup[subfigure]{oneside,margin={0.8cm,0cm}} %reset for figures (b) and (c)
\subfloat[]{\includegraphics[trim={0cm 0 0 0},clip,height=80pt]{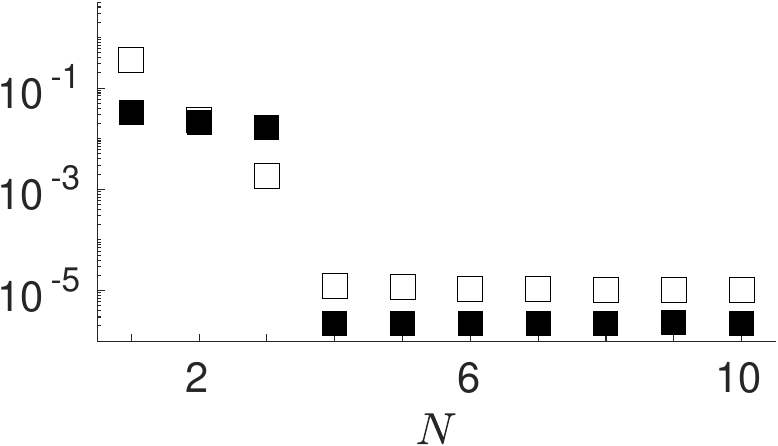}}
\caption{
Dependence of the residual $r$ on the number of terms $N$ retained in a given relation in the noiseless case. Black (white) squares represent data collected near the edge (in the middle) of the channel. The identified relations are (a) the energy equation, (b) the pressure equation,
%\textcolor{red}{(potentially with an additional term proportional to ${\bf u}^2$)}, 
and (c) the momentum equation.}
\label{fig:pareto}
\end{figure*}

Regression performed using the vector library $\mathcal{L}_1$ identifies a single relation representing momentum balance
\begin{align}\label{eq:NS_coeffs}
    c_1\partial_t {\bi u} + c_2({\bi u}\cdot \nabla){\bi u} + c_3\nabla p + c_4 \nabla^2 {\bi u}= 0. 
\end{align}
Table \ref{tab:ns} lists the surviving terms ${\bi f}_n$, the corresponding coefficients $c_n$, their uncertainties $s_n$, and the respective magnitudes $\chi_n=\|c_n{\bf q}_n\|$ of the terms which can be used to identify dominant balances in different regions. 
Here and below, uncertainties are estimated by repeating regression 128 times with a random subsample of half the spacetime domains and computing standard deviations.
For data from the edge of the channel, the Navier-Stokes equation with accurate coefficients, including the small viscosity, is identified for all noise levels. In particular, for noiseless data, the magnitude of the viscosity is identified correctly to several significant digits, even though the viscous term involves a second-order derivative, which is the highest in the equation. The corresponding sequence of residuals $r^{(N)}$ can be found in Figure \ref{fig:pareto}(c). 
 
For data from the middle of the channel, different special cases of equation \eqref{eq:NS_coeffs} are identified for different noise levels. For noise levels up to about 15\%, sparse regression still identifies the Navier-Stokes equation. For higher noise levels (up to 100\%), the Euler equation is identified instead, also with fairly accurate coefficients. For extreme levels of noise (up to 300\%), the inviscid Burgers equation is identified. Note that the sequence in which the terms in the momentum equation stop being recovered, as the noise level is increased, corresponds to their magnitudes $\chi_n$ in the middle of the channel.

No multi-term relations are found, for any data set, from the scalar library $\mathcal{L}_0$, suggesting it is incomplete. In order to identify any multi-term relations, this library was expanded by including %all linearly independent scalars quadratic in both ${\bi u}$ and $\nabla$, as well as a cubic in ${\bi u}$ term that should be present in a relation representing energy balance:
terms quartic in ${\bf u}$ and/or $\nabla$:
\begin{equation}
    \mathcal{L}_0'=\mathcal{L}_0\cup\{\nabla^2 ({\bi u}^2),\nabla \cdot \left[ ({\bi u} \cdot \nabla) {\bi u} \right],(\nabla \cdot {\bi u})^2,\nabla {\bi u}: \nabla {\bi u}^\top,\nabla {\bi u}:\nabla {\bi u},\nabla \cdot( {\bi u}^2{\bi u}),{\bi u}^4\}.
    \label{eq:augment_rank0}
\end{equation}
Whenever possible, the additional terms were written in conservative form to decrease numerical error associated with evaluation of higher-order derivatives in weak form. Two relations are identified from this expanded scalar library via sparsification:
\begin{align}
    & c_1 \partial_t E + c_2 \nabla \cdot({\bi u}E) + c_3 {\bi u} \cdot \nabla p + c_4\nabla^2E +c_5 \nabla {\bi u} : \nabla {\bi u} = 0 \label{eq:energy_coeffs},\\
 & c_6\nabla^2 p + c_7\nabla \cdot[({\bi u} \cdot \nabla) {\bi u}] +c_8 = 0, \label{eq:pp_coeffs}
\end{align}
where $E={\bf u}^2/2$ is the energy density.
The two relations are discovered robustly for both data sampled from the edge and the middle of the channel; however, the order in which they are found depends on the sampled region.
%The first one represents to the energy balance while the second one has a form similar to the pressure-Poisson equation; consequently we will refer to it as the pressure equation. 
The corresponding sequences of residuals $r^{(N)}$ can be found in Figure \ref{fig:pareto}(a-b), and the values of the coefficients $c_n$ for different noise levels are summarized in Tables \ref{table:energy_coeffs} and \ref{table:pp_coeffs}. 

For sufficiently low levels of noise (below about 10\%), relation \eqref{eq:energy_coeffs} corresponds to the well-known energy equation, again with fairly accurate coefficients, no matter which region of the flow the data comes from. The accuracy of the coefficients decreases somewhat as the level of noise is increased, as expected. For data from the middle of the channel, small terms such as $\nabla{\bf u}:\nabla{\bf u}$ start to disappear as the noise level is increased, similar to what we found for the momentum equation.   
Relation \eqref{eq:pp_coeffs} takes the form of the pressure-Poisson equation for moderately noisy data from both the middle and the edge of the channel, also with fairly accurate coefficients. The surprising observation is that, for noiseless data from the middle of the channel, sparse regression reliably identifies a small correction to the pressure-Poisson equation, a term proportional to unity. We will therefore refer to relation \eqref{eq:pp_coeffs} simply as the pressure equation.

Note that neither the energy equation nor the pressure-Poisson equation is an independent relation; both can be derived from the Navier-Stokes equation and the incompressibility condition. 
Further, note that SPIDER is superior to alternative approaches to equation inference in both versatility and accuracy: for instance, neither of the two scalar relations would be identified using SINDy, which assumes either $\partial_t{\bf u}$ or  $\partial_tp$ to be present. The energy equation has the form of an evolution equation but involves a temporal derivative of ${\bf u}^2$ rather than ${\bf u}$, while the pressure-Poisson equation is a constraint which involves no temporal derivatives at all. Finally, note that one can also heuristically identify the incompressibility condition, the pressure-Poisson equation and the energy equation from the right singular vectors of $Q$ corresponding to the three smallest singular values without pruning $\mathcal{L}'_0$. These singular vectors are dense, but only the coefficients associated with the corresponding dominant balances are O(1).

\begin{table*}
\centering
\subfloat[]{
\begin{tabular}{cccccc}
 & $\sigma$ & $\partial_t {\bi u}$ & $({\bi u}\cdot \nabla) {\bi u}$ & $\nabla p$ & $\nabla^2{\bi u}$ \\ \hline
$\bar{c}_n$ & 0\% & 0.999996 & 0.999998 & 1 & $-4.99996 \! \times \! 10^{-5}$ \\
& 50\% & 0.9928 & 0.995 & 1 & $-5.1\! \times \!10^{-5}$ \\ 
& 100\% & 0.9927 & 0.99 & 1 & $-5.2 \! \times\! 10^{-5}$ \\ \hline
$s_n$ & 0\% & $5\!\times\!10^{-8}$ & $3\!\times\!10^{-6}$ & $4\!\times\!10^{-6}$ & $3\!\times\!10^{-10}$ \\ 
 & 50\% & $1\!\times\!10^{-4}$ & $5\!\times\!10^{-3}$ & $8\!\times\!10^{-3}$ & $8\!\times\!10^{-7}$ \\
 & 100\% & $2 \!\times \! 10^{-4}$ & $1\!\times \! 10^{-2}$ & $2 \! \times \! 10^{-2}$ & $1 \! \times \! 10^{-6}$ \\ \hline
$\chi_n$ & 0\% & 0.79 & 1 & 0.50 & 0.45 \\
& 50\% & 0.79 & 1 & 0.51 & 0.46 \\
& 100\% & 0.82 & 1 & 0.51 & 0.47 \\ \hline
\end{tabular}
}
\\
%\hspace{20mm}
\vspace{2mm}
\subfloat[]
{
\begin{tabular}{*{7}{c}}
& $\sigma$ & $\partial_t {\bi u}$ &  $({\bi u}\cdot \nabla) {\bi u}$ & $\nabla p$ & $\nabla^2{\bi u}$ \\ \hline
$\bar{c}_n$ &0\% & 0.99986 & 1 & 0.99986 & $-5.003\!\times\!10^{-5}$ \\
& 50\% & 0.991 & 0.990 & 1 & 0  \\ 
& 100\% & 0.988 & 0.986 & 1 & 0  \\
& 300\% & 1 & 0.983 & 1 & 0 \\ \hline
$s_n$ & 0\% & $2\!\times\!10^{-7}$ & $1\!\times\!10^{-6}$ & $2\!\times\!10^{-5}$ & $2\!\times\!10^{-8}$ \\ 
& 50\% & $1\!\times\!10^{-4}$ & $1\!\times\!10^{-3}$ & $1\!\times\!10^{-2}$ & 0 \\
& 100\% & $2\!\times\!10^{-4}$ & $1\!\times\!10^{-3}$ & $2\!\times\!10^{-2}$ & 0  \\ 
& 300\% & $6\!\times\!10^{-4}$ & $5\!\times\!10^{-3}$ & 0 & 0  \\\hline
$\chi_n$ & 0\% & 0.99 & 1 & 0.07 & 0.006 \\
& 50\% & 0.99 & 1 & 0.07 & 0 \\
& 100\% & 0.99 & 1 & 0.07 & 0  \\
& 300\% & 1 & 0.99 & 0 & 0  \\ \hline
\end{tabular}
}
\caption{Coefficients of the momentum equation \eqref{eq:pp_coeffs} in the presence of varying levels of noise and data from (a) the edge of the channel and (b) the middle of the channel. The rows show the mean values of the coefficients $\bar{c}_n$ (normalized by the magnitude of the largest one), their uncertainties $s_n$, and the magnitudes of the terms $\chi_n$ (normalized by the magnitude of the largest term).
%Note that, at higher noise levels, the viscous term is discarded for the mid-channel data, where its magnitude is small compared to all the other terms.
}
\label{tab:ns}
\end{table*}

\begin{table}
\centering
\subfloat[]{
\begin{tabular}{ccccccc}  
\input{tables/energy_coeffs_edge.txt}
\end{tabular}
}\\
\vspace{1cm}
\subfloat[]{
\begin{tabular}{ccccccc}  
\input{tables/energy_coeffs_center.txt}
\end{tabular}
}
\caption{Coefficients of the energy equation \eqref{eq:energy_coeffs} in the presence of varying levels of noise and data from (a) the edge and (b) the middle of the channel. The quantities $\bar{c}_n$, $s_n$, and $\chi_n$ are defined in the caption of Table \ref{tab:ns}.}
\label{table:energy_coeffs}    
\end{table}

\begin{table*}
\centering
\subfloat[]{
\begin{tabular}{ccccc}
\input{tables/PP_coeffs_edge.txt}
\end{tabular}}
\hspace{1cm}
\subfloat[]{
\begin{tabular}{ccccc}
\input{tables/PP_coeffs_center.txt}
\end{tabular}}
\caption{Coefficients of the pressure equation \eqref{eq:pp_coeffs} in the presence of varying levels of noise and data from (a) the edge and (b) the middle of the channel. The quantities  $\bar{c}_n$, $s_n$, and $\chi_n$ are defined in the caption of Table \ref{tab:ns}.}
\label{table:pp_coeffs}
\end{table*}

\begin{figure*}
\centering
\captionsetup[subfigure]{oneside,margin={1.5cm,0cm}}
\subfloat[]{\includegraphics[width=0.4\textwidth]{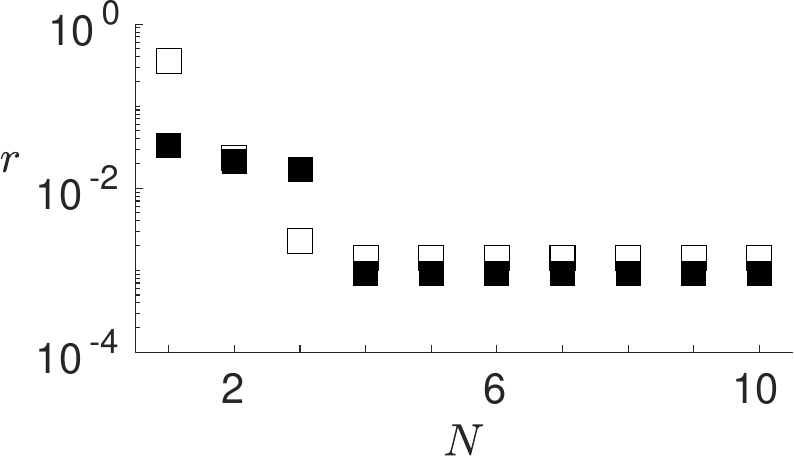}}
\hspace{1cm}
\subfloat[]{\includegraphics[width=0.4\textwidth]{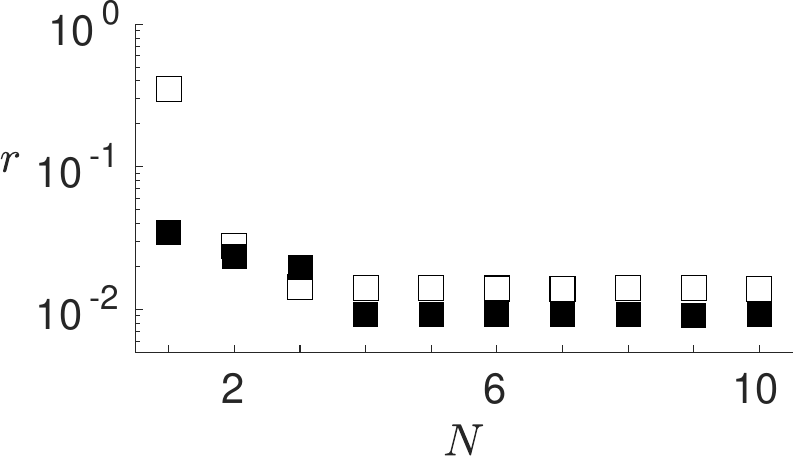}}
\\
\subfloat[]{\includegraphics[width=0.4\textwidth]{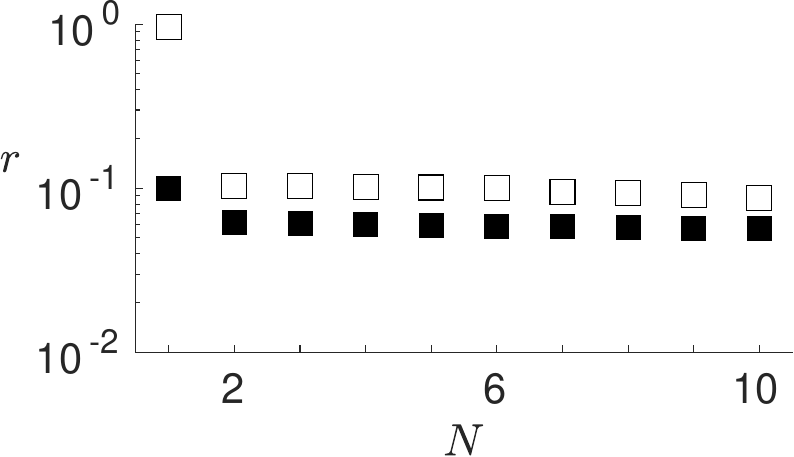}}
\hspace{1cm}
\subfloat[]{\includegraphics[width=0.4\textwidth]{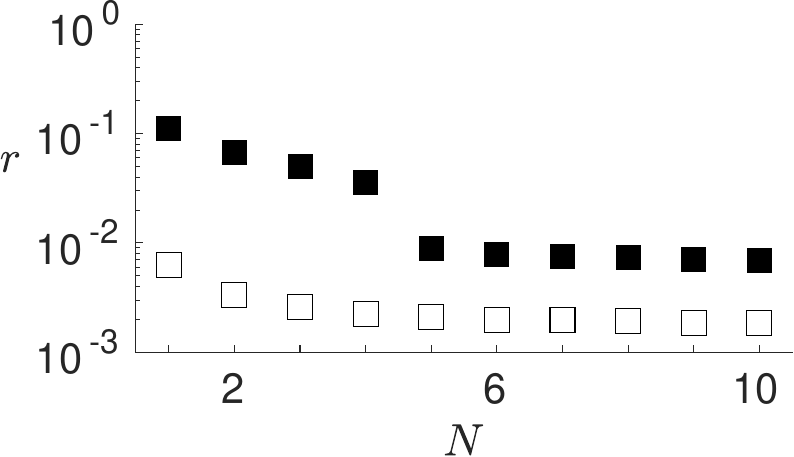}}
\caption{Dependence of the residual $r$ on the number of terms $N$ retained in the momentum equation for synthetic data with added noise for noise levels of (a) 10\% and (b) 100\%. The Navier-Stokes equation is identified in all the cases except for data from the middle of the channel with 100\% noise. In that case, the Euler equation is found instead. The residual for the pressure equation identified from data with 20\% noise (c) and for the energy equations identified from data with 10\% noise (d). Black (white) squares represent data collected near the edge (in the middle) of the channel.
}
\label{fig:pareto_noisy}
\end{figure*}

For both sampled regions, all libraries, and all noise levels, the residual $r$ asymptotes to a constant value for large $N$. In the noiseless case shown in Figure \ref{fig:pareto}, the asymptotic value of the residual is determined by the discretization of the data, both in the numerical simulations and in evaluating the integrals using quadratures, as shown by \citet{gurevich2019}. In the noisy case, shown in Figure \ref{fig:pareto_noisy}, the asymptotic value of the residual is instead determined by the level of noise and is higher than in the noiseless case, as expected. 
%However, qualitatively, the dependence of $r$ on the number of terms retained is the same in all of the cases.
It should be emphasized that physically meaningful relations can be identified in the presence of very high levels of noise, illustrating the robustness of SPIDER. 

%The critical number of terms $N$ depends both on the noise level and the dynamics displayed in the sampled data. The data with larger gradients (near the edge of the channel) make regression more robust to noise.

The power of the weak formulation compared with the strong form is vividly illustrated by other physical relations containing Laplacians as well. For instance, the vorticity equation can only be identified correctly in strong form for noise levels up to approximately 1\% as shown by \citet{rudy2017}. The pressure-Poisson equation \eqref{eq:pp_coeffs}, which also contains second-order derivatives, can be identified in weak form for noise levels up to 20\% (500\%) for data from the middle (edge) of the domain. The energy equation \eqref{eq:energy_coeffs} can be identified in weak form using the data from the edge of the channel with up to 10\% noise and using data from the middle of the channel with up to $3\%$ noise. The latter equation is more challenging to identify from noisy data due to the presence of the term $\nabla{\bf u} : \nabla {\bi u}$ which cannot be integrated by parts and was computed using second-order accurate finite differencing. However, it should be noted that a more sophisticated procedure for numerically differentiating the data would likely allow this equation to be identified even at higher noise levels. 

Note that, although we used the absolute residual $r$ in sparse regression, it is the magnitude of the relative residual $\eta$ that better quantifies the accuracy of the identified relations \citep{reinbold2021}. The relative residuals are indeed quite low: $4\!\times\!10^{-4}$ $(2\!\times\!10^{-2})$  for the energy equation, $2 \! \times \! 10^{-3}$ $(1\!\times\!10^{-4})$ for the two-term pressure-Poisson equation and $4\!\times\!10^{-5}$ $(5\!\times\!10^{-5})$ for the Navier-Stokes equation identified using noiseless data from the middle (edge) of the channel. The three-term pressure equation has a relative residual of $8 \! \times \! 10^{-4}$ in the middle of the channel, which is less than a half that of the pressure-Poisson equation.

\subsection{Learning boundary conditions}

SPIDER can also be used to discover boundary conditions. In this case, the rotational symmetry is partially broken: instead of rotations in all three spatial directions, the problem is only invariant with respect to rotations about the normal ${\bi n}$ to the boundary. {The reduced symmetry group describing boundary conditions is O(2).} The library of terms that transform as vectors near the boundary includes ${\bi n}$ in addition to ${\bi u}$ and $\nabla$. We exclude time derivatives, because these can be eliminated with the help of the bulk equations. We also exclude the dependence on $p$ to keep the library to a reasonable size (this dependence is trivial to restore). Retaining terms that contain each of ${\bi u}$ and $\nabla$ at most once {(${\bf n}$ has unit magnitude and is allowed to appear an arbitrary number of times)} yields a vector library
\begin{align}\label{eq:bclib}
    \mathcal{L}_1'=\{{\bi u},{\bi n},({\bi u} \cdot {\bi n}){\bi n},\nabla({\bi u} \cdot {\bi n}),({\bi n}\cdot \nabla){\bi u},{\bi n}(\nabla \cdot {\bi u})\}.
\end{align}
{Next, since they transform differently under rotation about the surface normal ${\bf n}$}, we separate the normal and tangential components by applying the projection operators $P_\perp={\bi n}{\bi n}$ and $P_\parallel=\mathbb{1}-{\bi n}{\bi n}$ to the library \eqref{eq:bclib}, where ${\bi n}{\bi n}$ represents the tensor product of the normal vectors. We prune all terms which have identically vanishing projections. Furthermore, we can also prune all terms involving $\nabla\cdot{\bi u}$, since we have already identified the continuity equation \eqref{eq:divu}. This results in two libraries for the boundary conditions:
\begin{align}\label{eq:bclib_proj}
    &\mathcal{L}_\parallel = \{ P_\parallel {\bi u}, P_\parallel\nabla({\bi u} \cdot {\bi n}), P_\parallel({\bi n}\cdot \nabla){\bi u}\},\\
    &\mathcal{L}_{\perp} = \{{\bi n} \cdot {\bi u},1,{\bi n} \cdot \nabla({\bi u} \cdot {\bi n}),{\bf n} \cdot ({\bi n}\cdot \nabla){\bi u}\},
\end{align}
{corresponding, respectively, to the vector and scalar irreducible representations of the symmetry group O(2).} 

Since the boundary conditions only hold on the solid walls $y=\pm 1$, each projection of the relation \eqref{eq:bclib} is integrated over rectangular $(2+1)$-dimensional domains $\Omega_k$ of size $H_x \times H_z \times H_t$ confined to one of the walls. Correspondingly, the weight functions ${\bi w}_j$ are constructed as products of three one-dimensional functions $\tilde{w}(s)$ where $s=\bar{x}$, $\bar{z}$, or $\bar{t}$. Note that the derivatives of the data with respect to the wall-normal ($y$) coordinate cannot be eliminated using integration by parts in this case; instead, we evaluate them directly using finite differences, although other alternatives could be used as well.
For all noise levels up to the maximum of $50\%$, valid single-term boundary conditions were always identified. Specifically, for the normal component, 
%two equivalent relations $P_\perp {\bi u} = P_\perp ({\bi u} \cdot {\bi n}){\bi n} = {\bi u} \cdot {\bi n} = 0$ are 
the relation ${\bi u} \cdot {\bi n} = 0$ is 
identified. For the tangential component, the relation $P_\parallel {\bi u} = 0$ is identified. These can be combined into the algebraically ``simplest'' boundary condition
\begin{align}\label{eq:bc}
    {\bi u}=0,
\end{align}
which is the well-known no-slip boundary condition.

%To summarize, SPIDER can successfully identify a complete, quantitatively accurate mathematical model of the fluid flow---the Navier-Stokes equation, the incompressibility condition, and the corresponding boundary conditions---using a tiny fraction of the available data, even when the data are extremely noisy. In fact, 
At sufficiently low noise levels, both the correct governing equations and the boundary conditions can be identified using only a single integration domain in the bulk and its projection onto the boundary, provided sufficiently many different weight functions are used. In particular, the no-slip boundary condition ${\bi u}=0$ and the incompressibility condition \eqref{eq:divu} are always correctly identified. 

\section{Discussion}
\label{sec:disc}

Physical constraints---the first key ingredient of SPIDER---play an essential role in the equation inference approach described here. The procedure used to construct the libraries of terms crucially relies on the irreducible representations of the symmetry group describing the physical problem. For the bulk equations, it is the orthogonal group O(3) describing rotations and reflections. In particular, the libraries $\mathcal{L}_0$ and $\mathcal{L}_1$ represent two of the irreducible representations of O(3)
corresponding to tensors of rank 0 and 1. Other irreducible representations of O(3) can be used to identify additional physical relations. For instance, the vorticity equation would require a library of antisymmetric rank-2 tensors, which are isomorphic to pseudovectors by contraction with $\varepsilon_{ijk}$. {For the boundary conditions, the symmetry group is O(2), representing rotations around the surface normal ${\bf n}$ and in-plane reflections. $\mathcal{L}_\parallel$ and $\mathcal{L}_\perp$ correspond to two different irreducible representations of O(2), vectors and scalars, respectively.}

Fluid flows have additional symmetries: translational invariance in space and time is responsible for all of the coefficients being constant. Furthermore, all governing equations should have Galilean invariance. We could have imposed this symmetry as a constraint from the start when constructing the libraries. This would have reduced the size of both scalar and vector library even further by constraining the temporal derivatives to only appear in the form of convective derivatives, e.g., as $\partial_tp+{\bf u}\cdot\nabla p$ in $\mathcal{L}_0$ and $\partial_t{\bf u}+{\bf u}\cdot\nabla{\bf u}$ in $\mathcal{L}_1$. Instead, we have let the data uncover this symmetry for us: inspection of the coefficients shows that both identified equations involving temporal derivatives acquire an explicitly Galilean-invariant form
\begin{align}
% & \nabla \cdot {\bi u} = 0 \nonumber \\
 &[\partial_t + {\bi u}\cdot \nabla]{\bi u} + \nabla p - \nu_1 \nabla^2 {\bi u}= 0, \nonumber\\%\label{eq:energy},\\
% & \nabla^2 p + \nabla \cdot[({\bi u} \cdot \nabla) {\bi u}] = 0 \label{eq:pp}\\
 & [\partial_t+{\bf u}\cdot\nabla]E + {\bi u} \cdot \nabla p - \nu_2 \nabla^2E + \nu_3 \nabla {\bi u} : \nabla {\bi u} = 0, \label{eq:Galilean}
\end{align}
with some positive coefficients $\nu_i$ after the incompressibility condition is applied.

Physics also dictates that all data not only transform in a particular manner under various symmetries---pressure as a scalar and velocity as a vector---but have appropriate dimensions or units. There is no need to explicitly enforce dimensional homogeneity of all the terms in the relations \eqref{eq:volterra}; this is accomplished by properly nondimensionalizing the terms ${\bf f}_n$ and treating the coefficients $c_n$ as dimensionless constants. However, the physical units determine the scales $S_i$ that play an essential role in nondimensionalization. This step is absolutely critical to the success of sparse regression, as the magnitudes of the coefficients $c_n$ are only meaningful once the terms ${\bf f}_n$ have been nondimensionalized using proper scales. In particular, it would be entirely unclear whether any single-term relation, such as the incompressibility condition, is appropriate without a proper scale to compare it with.

The weak formulation---the second key ingredient of our approach---imparts SPIDER with unprecedented robustness, allowing it identify correct physical relations from data with extreme levels of noise, making it indispensable for analyzing experimental data. Weak formulation also allows SPIDER to identify extremely subtle physical effects such as viscous stresses near the midplane of the flow where velocity gradients are small. Note that all of the coefficients in the energy and momentum balance equations \eqref{eq:Galilean} are very close to their true values of either unity or the viscosity $\nu_0 = 5 \times 10^{-5}$ used in the numerical simulations. 
Table \ref{tab:viscosity_coeffs} shows the deviation of the learned viscosity coefficients $\nu_i$ from the actual value. In particular, the viscosity $\nu_1$ appearing in the momentum equation is identified with remarkable precision, especially near the boundary, where the velocity gradients are large. On the contrary, the values of the viscosity $\nu_2$ and $\nu_3$, which appear in the energy equation, are substantially less accurate, which reflects the manner in which the energy dissipation term $\nabla{\bf u} : \nabla {\bi u}$ is computed. There is no way to move all of the derivatives contained in this term onto the weight function, so these derivatives must be calculated numerically. In this work, we used central finite differencing for all first-order derivatives that cannot be eliminated, incurring a substantial error quadratic in the grid spacing. To obtain $\nu_2$ and $\nu_3$ with higher accuracy from noiseless data, a higher-order differentiation scheme could, in principle, be employed.

\begin{table}[b]
    \centering
    \begin{tabular}{c| c c| c c |c c}
            & $\nu_1/\nu_0$ & $s_1/\nu_0$ &   $\nu_2/\nu_0$ & $s_2/\nu_0$ & $\nu_3/\nu_0$ & $s_3/\nu_0$ \\ \hline
    edge   & 0.999992 & $6 \times 10^{-6}$& 1.04 & 0.01 & 1.04 & 0.01  \\
    center & 1.0006  & $4 \times 10^{-4}$ & 0.988 & 0.004 & 0.972 & 0.04  
    \end{tabular}
    \caption{The mean values $\nu_i$ and uncertainties $s_i$ of the coefficients corresponding to viscosity in equations \eqref{eq:Galilean}, all normalized by the true viscosity $\nu_0 = 5 \times 10^{-5}$ of the dataset. The uncertainty is estimated by rerunning the regression $M/2=128$ times using a random sample of only half of the integration domains and taking the sample standard deviation of the resulting coefficient vectors.}
    \label{tab:viscosity_coeffs}
\end{table}

%In the center of the channel, the effect of viscosity is negligible due to the lack of small-scale structures, so the flow is described quite accurately by both the Navier-Stokes and the Euler equation. Indeed, in that region of the flow, the dominant balance is between the terms $\partial_t{\bi u}$ and $({\bi u}\cdot\nabla){\bi u}$, while the viscous term is more than two orders of magnitude smaller. The linear term $\alpha{\bi u}$ is similarly small. It is then not surprising that, at higher noise levels, the greedy algorithm discards both the viscous and the linear term, yielding the Euler equation.

Sparse regression is the third key ingredient of SPIDER, and our regression algorithm based on singular value decomposition of the feature matrix has several advantages compared with SINDy and its various alternatives. First of all, as mentioned previously, the important dominant balances can be identified by inspecting the smallest singular values even without performing sparsification. Furthermore, the magnitude of the smallest singular value can be used to determine, again without performing sparsification, whether the corresponding library contains any meaningful relations describing the data and, hence, whether it needs to be expanded.

For the vector library \eqref{eq:rank1}, very different dominant balances are found in the two sampled regions, as the magnitudes $\chi_n=\|c_n{\bf q}_n\|$ of different terms listed in Table \ref{tab:ns} illustrate. Near the boundary, all four terms in the Navier-Stokes equation are of comparable magnitude, so it is not surprising that the same relation is identified for all listed noise levels. For data from the middle of the channel, the dominant balance involves only the terms $\partial_t{\bf u}$ and ${\bf u}\cdot\nabla{\bf u}$. In comparison, the term $\nabla p$ is smaller by more than an order of magnitude, and the viscous term is smaller by more than four orders of magnitude. (That such a small viscous term can be identified---at noise levels that are as large as 15\%---is due mainly to the exceptional robustness of the weak formulation.) These large differences in the magnitudes of different terms explain the order in which the inviscid Burgers equation, the Euler equation, and the Navier-Stokes equation are identified as the noise magnitude is decreased. All three equations accurately describe the flow in the midplane of the channel and all three equations belong to the Pareto-optimal set generated by our greedy regression algorithm. Our choice of stopping criterion is one, but far from the only, way to choose between these three equations.

Let us comment on one unexpected result pointed out previously. As Figure \ref{fig:pareto}(b) illustrates, for noiseless data from the middle of the channel, {SPIDER fairly consistently identifies a spurious term $\alpha$ in the pressure equation:
\begin{equation}
\nabla^2 p + \nabla \cdot[({\bi u} \cdot \nabla) {\bi u}] + \alpha = 0.
\label{eq:modified_pp}
\end{equation}
Most commonly, this term is a small constant, as shown in Table \ref{table:pp_coeffs}, and its magnitude $\chi_n=O(10^{-3})$ is much less than unity. Including this extra term decreases the residual $r$ by a factor between 1.2 to 2 depending on the sample of integration domains, which does not always surpass the threshold $\gamma=1.3$ used in the greedy algorithm. However, equation \eqref{eq:modified_pp} is consistently identified as the most accurate three-term relation for four of the five data subsets described in the Appendix. For one subset ({\it center5}), the spurious term is instead consistently identified as a multiple of $E$, with a similarly small coefficient $c_n$ and magnitude $\chi_n$. Including either term does not produce a noticeable improvement in the residual for noiseless data from the edge of the channel ({\it edge1}), where the residual decreases by a much smaller factor of $1.02$. The presence of a spurious term likely reflects the limited accuracy of the numerical solution in the middle of the channel where the computational grid is the coarsest.}

\begin{figure*}
\centering
\subfloat[]{\includegraphics[width=0.47\textwidth]{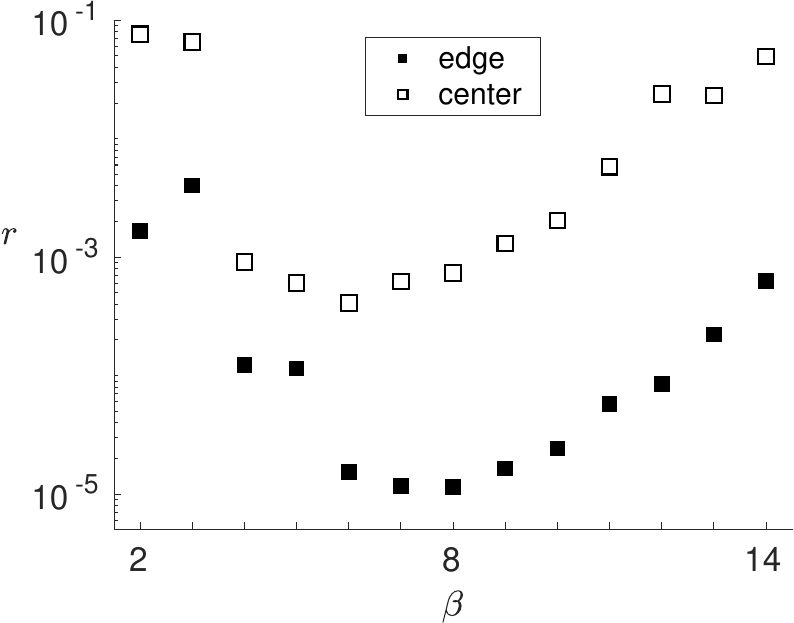}} \hfill
\subfloat[]{\includegraphics[width=0.47\textwidth]{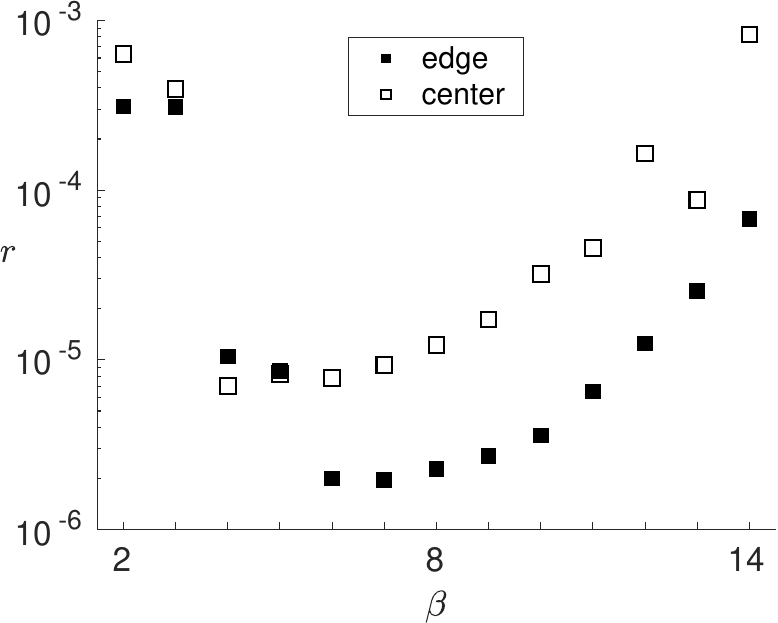}}
\caption{Dependence of residuals in (a) the pressure equation and (b) the momentum equation on the $\beta$ hyperparameter. The solid (dashed) curves correspond to data near the middle (edge) of the channel.
}
\label{fig:beta_dependence}
\end{figure*}

All the results presented here were obtained for the choice of the weight function exponent $\beta = 8$, which in Figure \ref{fig:beta_dependence} we find to roughly minimize the residuals of both the pressure-Poisson equation and the Navier-Stokes equation in both regions. Any choice in the range $6\leq \beta\leq 10$ yields comparable residuals. For uniform grids, it is advantageous to use higher values of $\beta$ (\citet{gurevich2019}); this improves the accuracy of the quadratures used in evaluating different library terms in weak from. The increase in the residual at higher values of $\beta$ is due to nonuniformity (in the wall-normal direction) of the computational grid on which the data is available.

{Finally, note that the approach presented here could be generalized to identify both governing equations and boundary conditions with parametric variation in space and/or time. Parametric variation can be easily detected by applying regression to subsets of data confined to small spatiotemporal volumes located at different positions. If the same functional relation is found but the coefficients differ, these variable coefficients could be replaced with a linear superposition of some basis functions and the regression repeated on the expanded library, as done by e.g., \citet{rudy2019}.}  

\section{Conclusion}
\label{sec:concl}

%\textcolor{blue}{In fact, even knowledge of the pressure field is not necessary to reconstruct the full set of governing equations for the present system. If we take only weight functions of the form ${\bi w}_j = \nabla \times {\bi A}_j$ for some set of vector-valued fields ${\bi A}_j$, then a version of the Navier-Stokes equation without the $\nabla p$ term can be reconstructed. Then, evaluating this modified equation with the usual set of weight functions reveals a nonzero residual term $e = -\nabla q$, which has vanishing curl \citep{reinbold2020}. The field $q$ thus defined represents an estimate of the pressure and can be treated as a new observable. However, such a procedure cannot be expected to identify latent variables in a general setting.}

To summarize, we have shown that a combination of very general physical constraints, weak formulation of PDEs, and sparse regression yields an extremely powerful model discovery tool, which we call SPIDER. It allows one to identify complete and easily interpretable quantitative mathematical models of continuum systems, {such as the highly turbulent fluid flow considered here}, from even very noisy data.
Moreover, SPIDER provides information about the relative importance of different physical effects in various regimes represented in the data.

The utility of the approach presented here is not limited to fluid dynamics. This same approach can be used to identify mathematical models of numerous high-dimensional, nonlinear, nonequilibrium systems that have defied traditional first-principles modeling approaches. Some examples include high energy density plasmas, as found inside the stars and the interior of fusion energy devices, and excitable media such as cardiac or intestinal muscle tissue and biological neural networks. Other interesting applications include active matter systems such as animal herds, bird flocks, insect swarms, fish schools, bacterial aggregates, self-propelled particles, and even collections of robots---these are formally discrete but may possess useful continuum models. Most active matter systems lack quantitative mathematical models while exhibiting interesting collective behaviors that could be better understood within the framework of such {continuum ``hydrodynamic''} models.

\section*{Acknowledgements}
%}
%\backsection[Funding]{
This material is based on work supported by the National Science Foundation under Grants No.~CMMI-1725587 and CMMI-2028454 and the Air Force Office of Scientific Research under grant FA9550-19-1-0005. D.R.G. was also supported by the NSF Graduate Research Fellowship Program.

\section*{Declaration of Interests}
The authors report no conflict of interest.

\bibliographystyle{jfm}
\bibliography{jfm}
%Use of the above commands will create a bibliography using the .bib file. Shown below is a bibliography built from individual items.

\appendix

\section{Sampled Data Locations}
The data used by SPIDER was obtained from the web cutout service of the Johns Hopkins Turbulence Database. The grid indices of the spacetime regions of the data are summarized in table \ref{tab:locations}.
Domains {\em center1} and {\em edge1} were used to generate all the reported results, while domains {\em center2} through  {\em center5} where used to investigate spurious terms in the pressure equation \eqref{eq:modified_pp}.

\begin{table}[]
    \centering
    \begin{tabular}{|c|c|c|c|c|}
    \hline
        domain ID & $I_x$ & $I_y$ & $I_z$ & $I_t$ \\
        \hline
        center1 & [1024,1088] & [256,320] & [750,814] & [2000,2064] \\
        center2 & [1,65] & [256,320] & [750,814] & [1000,1064] \\
        center3 & [1024,1088] & [256,320] & [750,814] & [500,564] \\
        center4 & [1524,1588] & [256,320] & [850,914] & [500,564] \\
        center5 & [1724,1788] & [256,320] & [850,914] & [500,564] \\
        edge1   & [1024,1088] & [1,65] & [768,832] & [2000,2065] \\
        \hline
    \end{tabular}
    \caption{The grid indices representing boundaries of the spacetime domains from which integration domains were sampled.}
    \label{tab:locations}
\end{table}

\end{document}

%% file: tables/energy_coeffs_edge.txt
 & $\sigma$  & $ \partial_t E $  & $ u_i \nabla_i p $  & $ \nabla_i(u_i E) $  & $ (\nabla_i u_j)(\nabla_i u_j) $  & $ \nabla^2 E $ \\ \hline
$\bar{c}_n$& 10\% & 1& 1.00 & 0.99418 & 0.0000404 & -0.0000473  \\ \hline 
$s_n$& 10\% & $ 5\!\times\! 10^{-3} $ & $ 2\!\times\! 10^{-2} $ & $ 3\!\times\! 10^{-5} $ & $ 5\!\times\! 10^{-7} $ & $ 7\!\times\! 10^{-7} $  \\ \hline 
$\chi_n$& 10\% & 0.69 & 0.27 & 0.75 &  1 & 0.76  \\ \hline 

%% file: tables/energy_coeffs_center.txt
& $\sigma$  & $ \partial_t E $  & $ {\bi u} \cdot \nabla p $  & $ \nabla \cdot ( {\bi u} E) $  & $  \nabla {\bi u} : \nabla {\bi u}  $  & $ \nabla^2 E $ \\ \hline
$\bar{c}_n$& 0\% & 0.99334 & 1& 0.993162 & 0.000048 & -0.0000494  \\ & 1\% & 0.99361 & 1& 0.993447 & 0.000050 & -0.0000493  \\ & 10\% & 0.9811 & 1& 0.98074 & 0& -0.000047  \\ \hline 
$s_n$& 0\% & $ 2\!\times\! 10^{-5} $ & $ 5\!\times\! 10^{-4} $ & $ 7\!\times\! 10^{-6} $ & $ 2\!\times\! 10^{-6} $ & $ 2\!\times\! 10^{-7} $  \\ & 1\% & $ 3\!\times\! 10^{-5} $ & $ 9\!\times\! 10^{-4} $ & $ 1\!\times\! 10^{-5} $ & $ 3\!\times\! 10^{-6} $ & $ 3\!\times\! 10^{-7} $  \\ & 10\% & $ 2\!\times\! 10^{-4} $ & $ 4\!\times\! 10^{-3} $ & $ 6\!\times\! 10^{-5} $ &  0 & $ 2\!\times\! 10^{-6} $  \\ \hline 
$\chi_n$& 0\% & 0.99 & 0.068 &  1 & 0.0011 & 0.0066  \\ & 1\% & 0.99 & 0.068 &  1 & 0.0012 & 0.0066  \\ & 10\% & 0.99 & 0.069 &  1 &  0 & 0.0064  \\ \hline 

%% file: tables/PP_coeffs_edge.txt
 & $\sigma$  & $ \nabla^2 p $  & $ \nabla_i \nabla_j (u_i u_j) $ \\ \hline
$\bar{c}_n$& 0\% & 0.99995 & 1 \\ & 100\% & 1.000 & 1 \\ & 500\% & 0.99 & 1 \\ \hline 
$s_n$& 0\% & $ 1\!\times\! 10^{-5} $ & $ 4\!\times\! 10^{-9} $  \\ & 100\% & $ 5\!\times\! 10^{-3} $ & $ 2\!\times\! 10^{-6} $  \\ & 500\% & $ 5\!\times\! 10^{-2} $ & $ 2\!\times\! 10^{-5} $  \\ \hline 
$\chi_n$& 0\% & 1.00 &  1  \\ & 100\% & 1.00 &  1  \\ & 500\% & 0.79 &  1  \\ \hline 

%% file: tables/PP_coeffs_center.txt
 & $\sigma$  & $ \nabla^2 p $  & $ \nabla_i \nabla_j (u_i u_j) $  & $ 1 $ \\ \hline
$\bar{c}_n$& 0\% & 1& 0.999789 & 0.00038  \\ & 10\% & 1& 0.9965 & 0 \\ & 20\% & 1& 0.9926 & 0 \\ \hline 
$s_n$& 0\% & $ 6\!\times\! 10^{-5} $ & $ 4\!\times\! 10^{-6} $ & $ 1\!\times\! 10^{-5} $  \\ & 10\% & $ 3\!\times\! 10^{-3} $ & $ 2\!\times\! 10^{-4} $ &  0  \\ & 20\% & $ 6\!\times\! 10^{-3} $ & $ 4\!\times\! 10^{-4} $ &  0  \\ \hline 
$\chi_n$& 0\% &  1 & 1.00 & 0.0016  \\ & 10\% & 1.00 &  1 &  0  \\ & 20\% & 0.99 &  1 &  0  \\ \hline 